\documentstyle[12pt]{article}
\begin{document}

\title{COSMICS: Cosmological Initial Conditions and Microwave Anisotropy Codes}

\author{Edmund Bertschinger\\
	Department of Physics, MIT\\
	Cambridge, MA 02139\\
 bertschinger@mit.edu}

\maketitle

\begin{abstract}
COSMICS is a package of fortran programs useful for computing transfer
functions and microwave background anisotropy for cosmological models,
and for generating gaussian random initial conditions for nonlinear
structure formation simulations of such models.  Four programs are
provided: {\bf linger\_con} and {\bf linger\_syn} integrate the linearized
equations of general relativity, matter, and radiation in conformal
Newtonian and synchronous gauge, respectively; {\bf deltat}
integrates the photon transfer functions computed by the linger
codes to produce photon anisotropy power spectra; and {\bf grafic}
tabulates normalized matter power spectra and produces constrained
or unconstrained samples of the matter density field.

Version 1.0 of COSMICS is available at {\bf http://arcturus.mit.edu/cosmics/}.
The current release gives fortran-77 programs that run on workstations and
vectorized supercomputers.  Unix makefiles are included that make it simple
to build and test the package. A future release will include portable parallel
versions of the linger codes using standard message-passing libraries.
\end{abstract}

\section{Introduction}

Theories of cosmic structure formation cannot be tested experimentally;
they must be simulated instead.  Numerical simulations of cosmic evolution
require three ingredients: assumptions about the cosmological model and
the matter and radiation content of the universe (e.g., a flat model
with cold dark matter, baryons, and a cosmological constant); a model for
the initial fluctuations (e.g., nearly scale-invariant fluctuations produced
by an early episode of inflation); and computer programs (and computers!)
for integrating the equations of motion.  The first two ingredients are at
the discretion of the simulator; the last one can be met, however, by
a standard package (computer not included).

COSMICS has been developed to provide some of the needed tools to the
cosmology simulation community.  It does not include programs for
simulating the complex nonlinear physics of structure formation; rather,
it generates accurate results of linear evolution.  For the microwave
background, excluding the Sunyaev-Zel'dovich effect and other minor
nonlinear corrections, this is sufficient for direct comparison with
observations.  For the matter distribution, on the other hand, the results
of COSMICS provide input to nonlinear evolution codes.

COSMICS is distributed over the world-wide web as a compressed tar file.
The package consists of 4 applications, a self-test, documentation (of
which this is part), and Unix makefiles.

A typical procedure for running COSMICS is the following.  First, one
runs {\tt linger} (either linger\_con or linger\_syn, depending on
details discussed below) to produce the data needed for matter transfer
functions or microwave background (CMB) anisotropy. Then one runs
{\tt grafic} to output the normalized matter power spectrum and, if desired,
to generate unconstrained or constrained (using the Hoffman-Ribak
algorithm) gaussian random density fields on a lattice (density, velocity,
and particle displacements).  {\tt Grafic} may obtain the matter transfer
function from {\tt Linger} or from an analytical fit, or it may use no
transfer function at all, resulting in scale-free fields.  The normalization
of the matter powe spectrum in {\tt grafic} may be specified either in
terms of the microwave background quadrupole moment $Q_{\rm rms-PS}$ or the
rms mass fluctuation $\sigma_8$; given one, the program computes the other.
Finally, if one is interested in accurate calculations of the CMB angular
power spectrum, {\tt linger} can be run at high resolution and the results
then processed by {\tt deltat}.  If one desires the angular spectrum to
high degree $l$, {\tt linger} runs are computationally demanding, requiring
tens of Cray C90 hours for $l\le2000$.  The expense comes because {\tt
linger} does the full calculation without significant approximations,
unlike most other codes in use today.  Most workers will not require this
accuracy; if they do, they may contact the author to see whether their
desired model has already been computed.  A future release of {\tt linger}
will include {\bf plinger} \cite{bodesc95,bodeaas186}, a parallel
implementation using message-passing that runs on a variety of
distributed-memory supercomputers.  A shared-memory version also exists
for the Cray C90.

COSMICS is available for academic use.  COSMICS users should notify me
by email to bertschinger@mit.edu, so that I can keep you informed of
upgrades and bug fixes.  Scientific publications using COSMICS should
acknowledge the author and the NSF under grant AST-9318185, which funded
the development of COSMICS.

\subsection{Building COSMICS}
\label{build}

COSMICS is easy to use.  First, get the compressed tar file
cosmics.tar.Z from \hfil\newline
\indent\indent{\tt http://arcturus.mit.edu/cosmics/}\hfil\newline
or by anonymous ftp to arcturus.mit.edu in directory Software (be
sure to set binary mode for the transfer).  Put it in a directory
with at least 10 MB of free space, then unpack it with\hfil\newline
\indent\indent{\tt uncompress cosmics.tar.Z; tar xf cosmics.tar }\hfil\newline
Go to the main directory, read the README file, and build the package
with make.  (First try make with no arguments, then select the desired
target.)  The makefiles are verified to work on a range of platforms
and operating systems (see the file Ported), but it is possible that
make will fail on your machine.   If it does, try {\tt make generic}.
If that fails, read Ported and try building COSMICS manually.  Then
send me email with a full description of what went wrong.  If you are
sufficiently skilled with Unix to solve the make problem yourself,
or you succeed in porting COSMICS to another machine, I would also
appreciate email so that I can incorporate these improvements into a
future release.  I will provide support for the ongoing use of this
package.

After the COSMICS codes are compiled, you can run a test with {\tt make test}.
This is a rather complete and lengthy test, requiring 27 MFlops-hours
(overnight on a typical workstation).  I could design a much shorter
test, but the main purpose is to acquaint you with some of the features
of COSMICS with realistic computations.  If you wish, you may try
{\tt linger\_con}, {\tt linger\_syn}, or {\tt grafic} out of the box ---
simply run the executables in subdirectory bin and answer the requests
for input parameters interactively.  After that (or before), read this
document to better understand the input and output, and what the COSMICS
programs are doing.

You can remove unwanted object files with {\tt make clean} in any of
the directories; doing this in the top directory will clean out all
of the subdirectories.  It will not remove the compiled binaries in
subdirectory {\tt bin}, or the files in {\tt test\_results}; these may be
removed with {\tt make realclean}.

\section{LINGER: Linear General Relativity}

{\tt Linger} integrates the coupled, linearized, Einstein, Boltzmann, and fluid
equations governing the evolution of scalar metric perturbations and photons
(both polarizations), neutrinos (both massless and, optionally, massive),
baryons, and cold dark matter in a perturbed flat Robertson-Walker universe.
In other words, it computes the linear evolution of fluctuations generated
in the early universe through the radiation-dominated era and recombination,
down to a small redshift input by the user.  The results are useful both for
calculations of the CMB anisotropy (with {\tt deltat}) and the linear power
spectrum of matter fluctuations (with {\tt grafic}).  {\tt Linger} provides
the link between the primeval fluctuations in the early universe and those
at late times (e.g., the present).  The {\tt linger} codes are described
in a preprint by Ma and Bertschinger \cite{mab1}.

Many other groups have written codes to integrate these equations (or a subset
of them): \cite{py70}--\cite{st94}; see \cite{hu} for a recent summary.
However, we believe that our treatment is the most accurate to date in its
treatment of the physics and the numerical integrations.  Our physics model
includes an accurate treatment of hydrogen recombination and the decoupling
of photons and baryons based on \cite{p68} with the addition of helium; helium
recombination using the Saha equation (this is adequate given the high
electron densities); a full treatment of Thomson scattering including two
photon polarizations and the full angular dependences of the scattering
cross section and distribution functions (see \cite{k95} for a complete
presentation of the theory); full computation of the gravitational sources
from all relevant particle species including all relativistic shear stresses
of photons and neutrinos; and full integration down to the final redshift
without use of any free-streaming approximation.  Our numerical methods
include a multipole expansion of the angular distribution of photons and
massless neutrinos to sufficiently high degree to accurately represent them
(up to $l=10000$ for late times and high spatial wavenumbers, when computing
CMB anisotropy; up to $l=100$ when computing matter transfer functions);
accurate sampling (with 128 points) of the momentum distribution of massive
neutrinos (and computation of the angular multipoles up to $l=50$); and
sufficiently fine sampling in the spatial wavenumber $k$ to give accurate
matter transfer functions and CMB anisotropy without any additional smoothing.

The aim of {\tt linger} is to produce results that are accurate to about 0.1\%.
This accuracy is, admittedly, somewhat artificial, since nonlinear effects
or other physics that is neglected by {\tt linger} may produce larger
differences.  (Research into this question is currently a focus of activity
for theoreticians investigating CMB anisotropy.)   However, I believe that
it is still worthwhile to solve the linear problem with such high precision.
Of course, all of this effort has a cost in the need for substantial computing
resources.  We discuss the requirements in section \ref{linger:usage}.
The user who wishes to can easily change {\tt linger} to be faster and
less accurate, by reducing the maximum multipole expansion orders {\tt lmax0}
and {\tt lmaxnu} set in fortran-77 {\tt parameter} statements in the code
(though be sure to search for occurrences in several subroutines).

The primary restrictions of the current release of {\tt linger} are: (1) it
assumes the unperturbed spacetime is flat, thereby excluding open or closed
models, and (2) only scalar (i.e., density) perturbations are included
(excluding vector and tensor perturbations, also known as gravitomagnetism
and gravitational radiation).  The second restriction is not a serious
limitation for computations of the matter fluctuation spectrum, although it
can lead to an underestimate of the large angular scale CMB anisotropy in
some cosmological models.  The first restriction, on the other hand, is more
serious given the interest among astronomers in testing open universe models
(despite the fact that such models lack a natural primeval fluctuation
spectrum).  However, {\tt linger} does allow for a cosmological constant,
so that $\Omega$ in matter may be less than unity.

{\tt Linger} comes in two versions, corresponding to two different gauge
choices for coordinates in the perturbed spacetime: synchronous gauge
({\tt linger\_syn}) and longitudinal or conformal Newtonian gauge
({\tt linger\_con}).  The latter case is equivalent to the so-called
gauge-invariant formalism.  (For a discussion of these and other gauges,
see refs. \cite{b80}--\cite{blh95}).  Although physically equivalent,
the output of the two codes is different.  Roughly speaking, the synchronous
case corresponds to using Lagrangian spatial coordinates that are fixed
with respect to the cold dark matter, while the conformal Newtonian case
corresponds to using Eulerian coordinates that (at late times) are (nearly)
fixed with respect to the microwave background.  See \cite{mab1} for the
transformation between the two sets of variables.

The two varieties of {\tt linger} are useful for different types of initial
conditions.  Isentropic (often inappropriately called adiabatic) initial
conditions, the type most naturally produced by cosmic inflation, may be
evolved equally well numerically in either gauge.  Many workers prefer the
conformal Newtonian gauge because the coordinates are minimally deformed
so that gauge variables are close to the quantities measured by Newtonian
observers.  Isocurvature fluctuations, which may be produced by first-order
phase transitions in the early universe, should be evolved in synchronous
gauge because they require fine-tuning in conformal Newtonian and other
gauges \cite{b80}.

Although the data output by the two versions of {\tt linger} differ because
of the gauges used, these differences do not affect their use because
physical observables are gauge-invariant.  {\tt Linger} output is used in
COSMICS for two purposes: computing the CMB angular power spectrum (in
{\tt deltat}) and computing and using the matter power spectrum (in {\tt
grafic}).  In the former case, the angular power spectrum $C_l$ is
gauge-invariant for $l>1$.  The monopole ($l=0$) is unobservable, while
the dipole ($l=1$) reflects the local motion of our galaxy and is
gauge-dependent simply because the coordinates of one gauge move relative
to those of another.  In synchronous gauge there is a very large $C_1$
(compared with the higher multipole moments) simply because the CMB
radiation has a large velocity ($\sim 500$ km s$^{-1}$) relative to the
rest frame defined by the matter --- the cold dark matter is, by definition,
at rest in synchronous coordinates.  In conformal Newtonian gauge, on
the other hand, the dipole moment is very small (comparable with the
higher multipoles) while the matter velocity is now nonzero.

The matter power spectrum used in {\tt grafic} is computed from the
gauge-invariant potential $\phi$ of conformal Newtonian gauge using the
Poisson equation:
\begin{equation}
\label{poisson}
\nabla^2\phi=-k^2\phi=4\pi G\bar\rho a^2\epsilon_m\ ,
\end{equation}
where $k$ is the spatial wavenumber and $a$ is the cosmic expansion
scale factor.  (This equation assumes that space curvature is negligible;
in section \ref{grafic2} below it is generalized to the case of open
models.)  On scales small compared with the Hubble distance, $\phi$
equals the Newtonian gravitational potential and $\epsilon_m$ is the net
matter density fluctuation; on larger scales they are the natural
generalized gauge-invariant variables defined by Bardeen \cite{b80}.
{\tt Linger\_con} outputs $\phi$; in {\tt linger\_syn} we output the
synchronous gauge metric variables plus the variable giving the exact
transformation to $\phi$.  So, either {\tt linger} code may be used, with
no difference whatsoever for {\tt grafic}, which automatically determines
the correct variables.  (The careful user should try both and compare them
as a test of speed and numerical precision.)

\subsection{Linger Usage}
\label{linger:usage}

After building the COSMICS package using {\tt make} in the main {\tt cosmics}
directory, the user should try running {\tt linger\_con} and {\tt linger\_syn}
interactively to gain familiarity with the input and output (the executables
are in subdirectory {\tt bin}).

\subsubsection{Linger Input}

{\tt Linger\_con} expects the following input:\hfil\newline
\indent\indent{\tt Omega\_b, Omega\_c, Omega\_v, Omega\_nu}\hfil\newline
\indent\indent{\tt H0, Tcmb, Y\_He, N\_nu(massive)}\hfil\newline
\indent\indent{\tt Bflag} [1 if full Boltzmann for CMB, 0 if lmax=100 for
    matter transfer functions]\hfil\newline
\indent\indent{\tt  kmax, nk, zend} [if Bflag=1] {\bf or}
               {\tt kmin, kmax, nk, zend} [if Bflag=0]\hfil\newline
Note that the fourth line of input requires 3 or 4 numbers depending on
whether {\tt Bflag} is set to 1 or 0.

{\tt Linger\_syn} expects the same input, except that one more parameter
is required at the end (the fifth line of input):\hfil\newline
\indent\indent{\tt ICflag} [=1,2,3,4 for isentropic or 3 kinds of isocurvature
fluctuations]\hfil\newline

These input parameters are mostly self-explanatory.  The {\tt Omega}'s are
the current (redshift zero) cosmic density parameter in baryons, cold
dark matter, vacuum energy (cosmological constant), and massive neutrinos,
respectively.  Currently, {\tt linger} is restricted to a flat background
spacetime, or $\Omega_b+\Omega_c+\Omega_v+\Omega_\nu=1$.  (Photons and
massless neutrinos also contribute to the energy density today, but their
effect is accounted for by a tiny shift in the Hubble constant from the
value input by the user: $H_{\rm 0,true}=H_0\left(1+\rho_r/\rho_m\right)^
{1/2}$, where $\rho_r$ is the present energy density of radiation (known
accurately through $T_{\rm cmb}$) and $\rho_m$ is the present energy density
of nonrelativistic matter.  The shift in $H_0$ is of no consequence except
for those users who wish to include relativistic particles in their
accounting of $\Omega$ and who are concerned with differences in $\Omega$
and $H_0$ at the level of .01\%.  {\tt Linger} uses the correct equations
internally given its redefinition of $H_0$.)

The next set of parameters are the Hubble constant $H_0$ in km s$^{-1}$
Mpc$^{-1}$, the microwave background temperature $T_{\rm cmb}$ in Kelvin,
the helium mass fraction of baryons $Y_{\rm He}$, and the number of
flavors of massive neutrinos $N_\nu$.  Standard parameters are suggested
by {\tt linger}.  Note that {\tt linger} fixes the total number of neutrino
flavors to be 3, so the number of massless flavors is $3-N_\nu$.  If
$N_\nu>1$, {\tt linger} assumes that all massive flavors have identical
mass.  Those who prefer a different pattern of neutrino masses should find
the modifications to {\tt linger} to be straightforward.

{\tt Bflag} is an important parameter directing {\tt linger} to run either
in an expensive mode with full resolution of the  angular power spectra
of photons and massless neutrinos, and with linearly spaced sampling in
spatial wavenumber $k$ ({\tt Bflag = 1}), or in a cheap mode with lower
angular resolution and logarithmically spaced sampling in $k$ ({\tt Bflag
= 0}).  The first mode is used for fully accurate CMB anisotropy calculations
(for {\tt deltat}); the second one is for matter transfer functions (for
{\tt grafic}).  The minimum and maximum spatial wavenumbers are $k_{\rm min}$
and $k_{\rm max}$, both measured in Mpc$^{-1}$.  ({\tt Linger} uses Mpc for
its units of length and time, not $h^{-1}$ Mpc.)  In the full Boltzmann case
({\tt Bflag = 1}), $k_{\rm min}=k_{\rm max}/nk$, where $nk$ is the number
of wavenumbers to compute.  In the reduced Boltzmann case ({\tt Bflag=0}),
the $nk$ wavenumbers are sampled logarithmically, starting at $k_{\rm min}$
and ending at $k_{\rm max}$.  The reason for these choices is that the
radiation transfer functions oscillate uniformly in $k$; sampling these
oscillations is needed for accurate integration of the angular power
spectrum without smoothing.  (However, Hu et al describe a smoothing
algorithm that works reasonably well with much coarser sampling \cite{hu};
perhaps a similar scheme might be incorporated into {\tt deltat} for
use with reduced-Boltzmann {\tt linger} runs.)  The matter transfer functions,
on the other hand, vary smoothly with $k$, and do not depend appreciably
on the high-order radiation multipole moments.  Finally, $z_{\rm end}$ is
the ending redshift of the computations.  {\tt Linger} outputs matter and
radiation transfer function data at this final redshift, with the specified
sampling in $k$.

The user with a typical workstation should not use {\tt Bflag = 1} except
if $k_{\rm max}\le0.1$ and/or $z_{\rm end}>0$.  The computing time for
each $k$-mode increases approximately linearly with $k$ because of the
need for the differential equation solver to sample the oscillations of
the photon and massless neutrino perturbations, whose frequency is
proportional to $k$.  Thus, most of the time is spent computing the
values near $k_{\rm max}$.  For flat models with $\Omega_v=0$, the CMB
anisotropy computed using $z_{\rm end}>0$ should agree rather well with
that computed with $z_{\rm end}=0$, aside from a compression of the
angular wavenumber $l$ because of the reduced distance to the cosmic
photosphere.  Experts can find other ways to further speed up the
CMB calculation \cite{hu}, albeit with a loss of accuracy.

The final parameter needed by {\tt linger\_syn}, {\tt ICflag}, is used
to set the type of initial conditions.  For {\tt ICflag = 1}, isentropic
initial conditions are selected, normalized so that the primeval
gauge-invariant potential $\psi(k)=-1$ for all $k$.  Isentropic initial
conditions correspond to primeval density fluctuations or, equivalently,
spacetime curvature fluctuations.  ($\psi$is one of the two scalar metric
variables of conformal Newtonian gauge; a gauge transformation is applied
to determine the metric variables in synchronous gauge \cite{mab1}.)
Exactly the same initial conditions are used by {\tt linger\_con}
(which is, however, restricted to isentropic initial conditions).  The
reason for the minus sign in $\psi$ is so that the density perturbations
in the nonrelativistic components will be positive from the Poisson
equation (\ref{poisson}); the amplitude is set arbitrarily to 1 so that
{\tt linger} calculates transfer functions normalized by the primeval
potential.  (This is the only physically sensible choice for isentropic
perturbations.)

For {\tt ICflag > 1}, isocurvature initial conditions are selected.  In
this case, the spacetime is initially unperturbed, but the ratios of
various matter and radiation components varies spatially.  {\tt ICflag
= 2} is the CDM entropy mode, for which the cold dark matter is assumed
to have spatial variations while the other components have much smaller
variations of the opposite sign because the initial conditions are set
when the universe is radiation-dominated \cite{be87}.  {\tt ICflag = 3}
is similar, except that here it is the baryons that vary initially,
compensated for by the other components.  Finally, {\tt ICflag = 4}
assumes that there is an additional component of static seed masses
such as primordial black holes; to a reasonable approximation this
also describes models with cosmics strings or textures.  In this case,
the other matter and radiation components are essentially unperturbed
initially, but the seeds provide a source term in the Einstein equations.
In all three isocurvature cases, the initial conditions are set so that
the density fluctuation in the spatially varying component is $\delta(k)=1$
for all $k$.

Typical values of {\tt  kmax, nk, zend} for {\tt Bflag = 1} (full Boltzmann)
runs are 0.5, 5000, 0.  These parameters yield integration errors less
than 0.15\% in the photon anisotropy spectrum $C_l$ up to $l_{\rm max}=3000$.
Such a run requires about 80 Cray C90 hours.  If accurate results are desired
for smaller $l_{\rm max}$, {\tt kmax} and {\tt nk} may be reduced
proportionally (keeping the ratio fixed).  Running {\tt linger} for
$l_{\rm max}=1000$ requires only about 10 C90 hours.

For matter transfer function runs ({\tt Bflag = 0}), {\tt linger} should
be run with input parameters set to {\tt kmin=1.e-5, kmax=10, nk=61} (or
121, for high accuracy), {\tt zend=0} or the desired starting redshift
for a nonlinear simulation ({\tt grafic} will automatically compute the
appropriate starting redshift and adjust the fluctuations back in time
if {\tt linger} is run with {\tt zend = 0}, so this is a safe choice).
The range of $k$ is set to ensure adequate sampling for computing the
CMB quadrupole moment (requiring small $k_{\rm min}$) and of the matter
transfer function at short wavelengths (requiring large $k_{\rm max}$).
{\tt Grafic} will extrapolate the transfer function beyond $k_{\rm max}$
if necessary.  It prints out a warning message when it does so; this
may generally be ignored if $k_{\rm max}\gg 1.0$ Mpc$^{-1}$ (so that the
transfer function is well-approximated by a power law).  With 61 points
for $k$ ranging from $10^{-5}$ to 10 Mpc$^{-1}$, {\tt linger\_con} requires
about 20 Mflops-hours and {\tt linger\_syn} about 15.

\subsubsection{Linger Output}

{\tt Linger} produces no standard output after the parameters are
entered; all subsequent output goes to two disk files, {\tt linger.dat}
and {\tt lingerg.dat}.  The first one gives, as a function of $k$ at
redshift $z_{\rm end}$, the metric variables and the density, velocity
divergence, and shear stress perturbations in all the components (except
that no shear stress is output for CDM and baryons, since they are, at the
final redshift, essentially perfect fluids with vanishing shear stress).
{\tt Linger.dat} is an ascii file with two header lines giving the input
parameters, followed by $nk$ lines giving the perturbation variables.
It is written as follows:
\hfil\newline{\tt
\indent\indent write(10,'(4(1pe12.6,1x))') Omega\_b,Omega\_c,Omega\_v,Omega\_nu
    \hfil\newline
\indent\indent write(10,'(3(1pe12.6,1x),3(i2,2x))')
H0,Tcmb,Y\_He,3-N\_nu,N\_nu,
    \hfil\newline
\indent\indent \& \indent ICflag\hfil\newline
\indent\indent do ik=1,nk\hfil\newline
\indent\indent\indent write(10,'(i7,1x,19(1pe11.4,1x))') ik,ak,a,tau,psi,phi,
    \hfil\newline
\indent\indent \& \indent
deltac,deltab,deltag,deltar,deltan,thetac,thetab,thetag,
    \hfil\newline
\indent\indent \& \indent thetar,thetan,shearg,shearr,shearn,econ\hfil\newline
\indent\indent end do\hfil\newline
}
{\tt ICflag} is set to 0 for {\tt linger\_con}, allowing one to determine
from the files which of the two codes was used for isentropic initial
conditions.  The other parameters are as follows:
{\tt ak} is the wavenumber in 1/Mpc, {\tt a = 1/zend - 1} is the
final expansion scale factor, {\tt tau} is the final conformal time in Mpc
(conformal time is related to proper time by $d\tau=dt/a$), {\tt psi} and
{\tt phi} are the metric perturbation variables (for {\tt linger\_con};
substitute {\tt ahdot} and {\tt eta} of \cite{mab1} in case of
{\tt linger\_syn}).  The {\tt delta}'s give the density fluctuations in
CDM (c), baryons (b), photons (g), massless neutrinos (r), and massive
neutrinos (n).  The {\tt theta}'s give the velocity divergence fluctuations
in the same components (except in the case of {\tt linger\_syn}, where
{\tt thetac} is replaced with the gauge transformation variable
{\tt phi-eta}, so that the gauge-invariant variable {\tt phi} of
conformal Newtonian gauge may be computed from the metric perturbation
variable {\tt eta} of synchronous gauge).  The {\tt shear}'s give the
anisotropic stress variable $\sigma$ of \cite{mab1}; it is related to
$\Pi$ of Kodama and Sasaki \cite{ks1} by $\sigma=2\Pi P/3(\rho+P)$ for
a component with mean density $\rho$ and mean pressure $P$.  Finally,
{\tt econ} is an energy conservation check computed using the 0-0
(energy constraint) Einstein equation; it gives a measure of the relative
accuracy of the numerical results.

{\tt Lingerg.dat} is an unformatted (binary) file containing the photon
intensity and polarization transfer functions.  It is written as follows:
\hfil\newline{\tt
\indent\indent write(11)
Omega\_b,Omega\_c,Omega\_v,Omega\_nu,H0,Tcmb,Y\_He,3-N\_nu,N\_nu,
    \hfil\newline
\indent\indent \& \indent ICflag\hfil\newline
\indent\indent write(11) Bflag\hfil\newline
\indent\indent write(11) kmin,kmax,nk,zend,tau\hfil\newline
\indent\indent do ik=1,nk\hfil\newline
\indent\indent\indent write(11) ik,ak,tau,lmax \hfil\newline
\indent\indent\indent write(11) (DeltaI\_l(k),l=0,lmax) \hfil\newline
\indent\indent\indent write(11) (DeltaQ\_l(k),l=0,lmax) \hfil\newline
\indent\indent end do\hfil\newline
}
Here, {\tt DeltaI\_l} is the perturbation in the photon temperature for
the lth multipole moment; {\tt DeltaQ\_l} is the perturbation in the
polarization.  (These two quantities are $1/4$ the perturbations in
the $I$ and $Q$ Stokes parameters; the factor of 4 providing the conversion
from intensity to temperature fluctuations.)  See \cite{mab1} for the
exact definitions (though note that {\tt DeltaI\_l} and {\tt DeltaQ\_l}
are written there as ${1\over4}F_{\gamma\,l}=\Delta_l$ and ${1\over4}
G_{\gamma\,l}$, respectively).

Because {\tt lingerg.dat} is unformatted, it cannot (usually) be read
on machines different from the one where it was created.  In a future
release of COSMICS, routines will be provided giving the conversion
of {\tt lingerg.dat} to and from a portable binary scientific data format
based on the NCSA HDF standard \cite{hdf}.

The results in {\tt linger.dat} are used by {\tt grafic}; the results in
{\tt lingerg.dat} are used by {\tt deltat}.  These codes are discussed next.

\section{DELTAT: Evaluate CMB Anisotropy Spectrum}

The photon temperature angular power spectrum is given by an integral
over spatial wavenumbers as \cite{mab1}
\begin{equation}
\label{cls}
  C_l=4\pi\int d^3k\,P(k)\,\Delta_l^2(k,\tau)\ ,
\end{equation}
where $P(k)$ is the power spectrum of the primeval potential $\psi$ for
isentropic initial conditions, or of the fluctuating density component for
isocurvature initial conditions, and $\Delta_l(k,\tau)$ is the total
temperature fluctuation (summed over polarizations) at the conformal time
$\tau$ corresponding to the ending redshift of {\tt linger}.  The angular
power spectrum is related to the angular correlation function $C(\theta)$
(here $\theta$ is an angle, not $\vec\nabla\cdot\vec v\,$!) by
\begin{equation}
\label{angcor}
  C(\theta)=\sum_l {2l+1\over 4\pi} C_l\,P_l(\cos\theta)\ ,
\end{equation}
where $P_l$ is a Legendre polynomial.  {\tt Deltat} performs the numerical
quadrature in equation (\ref{cls}) using Romberg integration of a continuous
$\Delta_l(k)$ interpolated from the values stored in {\tt lingerg.dat}
using cubic splines.  The radiation transfer functions undergo damped
oscillations with slowly-varying amplitude and phase at fixed $\tau$:
$\Delta_l(k)=A_l(k)j_l(k\tau+\varphi_l(k))$ ($j_l$ is a spherical Bessel
function).  It is difficult to determine or fit $A_l(k)$ and $\varphi_l(k)$,
so instead we use the numerically computed values of $\Delta_l(k)$,
interpolated with a spline.  The point of this discussion is that
$\Delta_l(k)$ oscillates rapidly, with a period of about $2\pi/\tau$;
$\tau\approx 2c/H_0=6000\,h^{-1}$ Mpc today.  Thus, our interpolation
method requires sampling in $k$ with $\Delta k \approx \tau^{-1}$ or better.
Although this sampling requirement is stringent, the advantage of our method
is that the spline provides an excellent fit and the Romberg integration
provides an extremely precise numerical quadrature.  Even with this
precision, {\tt Deltat} runs much more quickly than {\tt linger} for
a full Boltzmann run.

The user of {\tt deltat} also needs to note that the integral in equation
(\ref{cls}) must be carried to $k_{\rm max}\approx 0.5 l_{\rm max}/\tau$
to get all significant contributions (to a level of 0.15\%); for higher $k$
the radiation transfer functions are negligible.  The user can experiment
with these parameters.

{\tt Deltat} is easy to run.  Prompted by the program, the user must
input the following: $l_{\rm max}$ and $n$; $l_{\rm save}$; and the
{\tt lingerg.dat} filename from {\tt linger} (the name should be changed
to avoid overwriting the file by later {\tt linger} runs).  The first
parameter requested by {\tt deltat} is simply the maximum $l$ to compute
$C_l$; {\tt deltat} uses the minimum of this value and the $l_{\rm max}$
for the radiation transfer functions in {\tt lingerg.dat}.  The parameter
$n$ is related to the logarithmic slope of the primeval power spectrum $P(k)$
in equation (\ref{cls}): $P(k)\propto k^{n-4}$.  The offset of 4 is due to
the unfortunate usage in cosmology of $n$ for the logarithmic slope of
the power spectrum of the gauge-invariant total density fluctuation
$\epsilon_m$ and not the physically relevant quantities.  For the standard
scale-invariant Harrison-Zel'dovich spectrum, $n=1$ for both isentropic
and isocurvature fluctuations.  The third numerical parameter,
$l_{\rm save}$, is simply a flag instructing the program to extract
$\Delta_l(k)$ for $l=l_{\rm save}$ from {\tt lingerg.dat} and write it
to an ascii file.  The user could do this by writing a small program, but
often it is useful to plot one of the radiation transfer functions as a
sanity check when one is in no mood to write such a program.

The output of {\tt Deltat} is equally simple: an ascii file {\tt deltat.dat}
containing 3 header lines (the same 2 as {\tt linger.dat}, plus an extra header
line giving $n$), followed by $l$ and the net power $l(l+1)C_l$ in the left
and right columns, respectively.  Additionally, if $0\le l_{\rm save}\le
l_{\rm max}$, an ascii file {\tt deltal.dat} is created containing one
header line with $l_{\rm save}$ followed by $k$ and $\Delta_l(k)$ in the
left and right columns, respectively.

\section{GRAFIC: Gaussian Random Field Initial Conditions}

{\tt Grafic} normalizes the power spectrum of matter density fluctuations
(either derived from {\tt linger.dat}, or from a standard parameter fit
to the CDM transfer function \cite{bbks}), and generates initial conditions
needed for nonlinear cosmic structure formation simulations.   It produces
the density fluctuation field $\epsilon_m(\vec x\,)$ (that is, $\delta\rho/
\rho$) in comoving coordinates as a gaussian random field with the
appropriate power spectrum.

Constraints may be imposed (such as the presence of a specified overdensity
of the smoothed density field) by providing them in a subroutine;
the Hoffman-Ribak algorithm \cite{hr} is used to correctly sample the
constrained action.  Our implementation method is described in a paper
giving a detailed presentation of the theory of constrained gaussian random
fields \cite{vdw}.  The HR algorithm has also been implemented by Ganon
and Hoffman \cite{gh}; their implementation is restricted to constraints
that may be specified at lattice points (as opposed to the arbitrary linear
constraints allowed by {\tt grafic}), but it is faster than {\tt grafic}
for more than a few constraints.  Note that {\tt grafic} is an exact
method, unlike the iterative heat bath algorithm developed earlier by
the author \cite{b87}, so that it is fast for up to tens of constraints.
The main limitation is on the memory required to store the constraint matrix.

{\tt Grafic} outputs both the density field and the initial positions and
velocities of particles displaced from the lattice to produce that density
field.  The former object is useful for initializing cosmological gas
dynamics solvers, while the latter quantities are needed for cosmological
$N$-body simulations.  They are related to each other using the Zel'dovich
approximation \cite{zel70}:
\begin{equation}
\label{zeld}
  \vec x\,(\vec q,\tau)=\vec q+D_+(\tau)\vec d\,(\vec q\,)\ ,\ \
  \vec v\,(\vec q,\tau)=\dot D_+(\tau)\vec d\,(\vec q\,)\ ;\quad
  \vec\nabla\cdot\vec d=-D_+^{-1}\epsilon_m(\vec q,\tau)\ .
\end{equation}
Here $\vec q$ is a Lagrangian coordinate corresponding to the unperturbed
comoving position of a mass element; {\tt grafic} takes these positions
to be on a regular Cartesian lattice with periodic boundary conditions.
The perturbed comoving positions are $\vec x$; the perturbations to position
grow in proportion with the cosmic growth factor $D_+(\tau)$, which depends
on the cosmological model.  The displacement field $\vec d\,(\vec q\,)$
is obtained by calculating the inverse Laplacian of the linear density field
using a fast fourier transform.  The approximation comes in the third of
equations (\ref{zeld}), which neglects terms $O(\epsilon_m^2)$.  {\tt Grafic}
automatically selects the output redshift high enough so that the maximum
density fluctuation at any lattice point has amplitude 1; for $64^3$ or
more  points this means that the rms density perturbation is typically less
than 0.2.  The proper peculiar velocity $\vec v$ follows straightforwardly.
{\tt Grafic} includes subroutines that compute $D_+(\tau)$, $\dot D_+(\tau)$,
$a(\tau)$, etc., for general Friedmann-Robertson-Walker models with matter,
vacuum energy, and curvature.

\subsection{GRAFIC Usage}

{\tt Grafic} can be used as is if one is interested only in outputting
the linear power spectrum of matter fluctuations and normalizing it to
the CMB quadrupole and/or $\sigma_8$.  However, if one wants to output
density, position, and velocity fields on a lattice, then one must
specify the lattice size and spacing and the constraints, if any.
This is done through an include file, {\tt grafic/grafic.inc}, and a
constraints subroutine, {\tt grafic/constr.f}, that the user must edit
before building {\tt grafic}.  There is a README file describing the process.

{\tt Grafic} input is slightly complicated owing to its flexibility.
It should be run interactively first for practice.  The first item
requested by {\tt grafic} is a flag ({\tt Tflag}) specifying the type
of matter transfer function to be used: {\tt Tflag = 1} to use the
{\tt linger.dat} previously computed by {\tt linger\_con} or {\tt
linger\_syn}; {\tt Tflag = 2} to use instead an analytic transfer
function fit to CDM models by Bardeen et al \cite{bbks}; or {\tt Tflag
= 3} if no transfer function should be applied at all to the underlying
power-law spectrum.  The latter case is appropriate for scale-free
simulations.  In the second case, it is straightforward to modify
the power spectrum routine {\tt power.f} if the user wishes to use
some other analytical form for the matter transfer function.

\subsubsection{GRAFIC Input 1: Using linger.dat for transfer functions}
\label{grafic1}

Each of the three cases has slightly different input after setting
{\tt Tflag}.  We shall begin with {\tt Tflag = 1}.  In this case, the
user inputs the name of the {\tt linger.dat} filename produced by
{\tt linger}.  (Its name should be changed to avoid overwriting
by subsequent {\tt linger} runs.)  From the {\tt linger} header
information, {\tt grafic} automatically determines the cosmological
parameters.  It then asks the user to enter the long-wave spectral
index $n$ (the same parameter used by {\tt deltat}; $n=1$ for the
scale-invariant Harrison-Zel'dovich spectrum).  Next, it requests the
desired normalization at redshift zero ($a=1$), either $Q_{\rm rms-PS}$
in $\mu\,$K if the user wishes to use a COBE normalization, or
$\sigma_8$ if the user prefers the conventional normalization on
galaxy cluster scales.  To distinguish them, a negative value should
be used for $\sigma_8$; {\tt grafic} then takes the absolute value.
These normalization quantities are defined as follows:
\begin{equation}
\label{norm}
  Q_{\rm rms-PS}\equiv T_0\left(5C_2\over4\pi\right)^{1/2}\ ,\quad
  \sigma_8\equiv\int d^3k\,P_\epsilon(k)\left[3j_1(kR_8)/(kR_8)\right]^2\ .
\end{equation}
Here, $T_0$ is the present-day microwave background temperature;
$C_2$ is the $l=2$ component of the angular power spectrum computed
by {\tt grafic} using equation (\ref{cls}), with $\Delta_2(k)={\tt
shearg/2}$ coming from the photon shear stress in {\tt linger.dat};
$P_\epsilon(k)$ is the matter density fluctuation power spectrum,
related to the primeval spectrum $P(k)$ of equation (\ref{cls}) via the
Poisson equation (\ref{poisson}); $j_1$ is the spherical Bessel function;
and $R_8=8$ $h^{-1}$ Mpc is the standard radius for computing $\sigma_8$.
The term inside brackets in the integral for $\sigma_8$ is the window
function for a spherical tophat, so that $\sigma_8$ is the rms density
fluctuation in a sphere of radius $R_8$.  Whichever way the user chooses
to normalize the power spectrum, {\tt grafic} quickly computes the other
quantity appropriate for this normalization from equation (\ref{norm}).
See the file {\tt grafic/accuracy\_considerations} for comments about the
accuracy of $Q_{\rm rms-PS}$.

The normalization quantities are evaluated at $a=1$ ($\tau=\tau_0$).
If {\tt linger} was evolved to {\tt zend} $>0$ ($\tau<\tau_0$), {\tt grafic}
corrects $\Delta_2(k)$ and $\epsilon_m(k)$ using linear theory in a
Friedmann universe (with matter and, possibly, vacuum energy, but
negligible radiation):
\begin{equation}
\label{z0corr}
  \Delta_2(k,\tau_0)=\Delta_2(k,\tau)+2\int_\tau^{\tau_0}
    j_2(k(\tau_0-\tau))\,\dot\phi(k,\tau)\,d\tau\ ,\quad
  \epsilon_m(k,\tau_0)={D_+(\tau_0)\over D_+(\tau)}\,\epsilon_m(k,\tau)\ .
\end{equation}
In the integral for $\Delta_2$, $\dot\phi$ is computed using the evolution
of the potential in a Friedmann universe, $\phi(k,\tau)\propto D_+(\tau)/
a(\tau)$.  These time-dependent quantities are computed accurately by
{\tt grafic}.

After the normalization is completed, {\tt grafic} optionally will output
to file {\tt power.dat} the matter power spectrum $P_\epsilon(k)$ at
$a=1$.  The user is prompted to enter $k_{\rm min}$ and $k_{\rm max}$
for this output; if either one is zero or negative, {\tt grafic} skips
this output.

Next, {\tt grafic} requests parameters used in constructing realizations
of the density field and particle positions and velocities.  The user
must enter {\tt dx}, {\tt epsilon}, and {\tt etat}.  The first quantity
is the lattice spacing in comoving Mpc; {\tt epsilon} is the desired softening
distance (in comoving Mpc) for subgrid-resolution simulation programs
such as particle-particle/particle-mesh (p3m, not included in COSMICS,
but part of a package of $N$-body solvers to be released by the author
in the future); and {\tt etat} is a parameter used by the author in p3m
to select the code timestep.  These parameters, among others, are output
by {\tt grafic} in header records for the particle output file.  Users may
wish to tailor the input of {\tt grafic} here for their needs, so that they
can write the output files in their own favorite formats.

{\tt Grafic} then requests a 9-digit random number seed to initialize
its pseudorandom number generator.  The random number routines, in
{\tt random8.f}, are based on a subtract-with-borrow lagged Fibonacci
generator with base $2^{32}-5$ and period $10^{414}$ \cite{mz}, shuffled by a
completely independent generator.  Although relatively expensive,
{\tt random8} produces psuedorandom numbers with a uniform distribution
and no detectable serial correlations (despite many attempts by the author
to find them when he got curious results due to bugs elsewhere!).

Finally, {\tt grafic} requests the user to enter a flag {\tt ido},
determining whether it will compute an unconstrained realization of
a gaussian random field ({\tt ido = 1}), the mean field of constraints
({\tt ido = 2}, this is not a noise field at all, but rather the ensemble
average of constrained noisy fields), or a realization of the constrained
random field ({\tt ido = 3}).  The last two options require that the
user set the constraints appropriately by editing {\tt constr.f}.

\subsubsection{GRAFIC Input 2: Using an analytical transfer function}
\label{grafic2}

If the user prefers to normalize the matter power spectrum for a
physical model without {\tt linger} output (this is useful for
exploratory studies, or for nonflat models), {\tt Tflag = 2} is the
appropriate choice.  In this case, {\tt grafic} cannot determine the
desired cosmological parameters from {\tt linger.dat}, so instead the
user is immediately prompted for $\Omega_m$, $\Omega_v$, and $H_0$
(in km s$^{-1}$ Mpc$^{-1}$; ignore the message about $H_0=1$).  This
case is not limited to a flat ($\Omega_m+\Omega_v=1$) cosmology, since
{\tt grafic} uses linear theory results valid in any Friedmann universe,
including open ones.  (Closed universes are currently out of fashion;
the correct treatment in this case is complicated slightly by the
fact that the spatial frequency becomes a discrete variable.  {\tt Grafic}
is not set up for this.)

After the user enters the cosmological parameters, {\tt grafic} requests
the long-wave spectral index $n$ and the normalization constant
($Q_{\rm rms-PS}$ or $-\sigma_8$), as discussed in section \ref{grafic1}.
({\tt Grafic} sets $T_0=2.726$ K; when {\tt linger.dat} is used, $T_0$
is read from the header.)  At this point onwards, the input for {\tt grafic}
is the same as in the first case (sec. \ref{grafic1}).  However, there are
some important issues to consider for an open universe, which is allowed here
but not at present with {\tt linger}.

{\tt Grafic} computes the CMB normalization using the first of
equations (\ref{z0corr}), except that $\tau$ is replaced by the conformal
time at recombination (setting it at $a=1200$) and the Sachs-Wolfe terms
are added for the intrinsic anisotropy at the cosmic photosphere in the
instantaneous-recombination approximation \cite{sw67}.  {\tt Grafic}
also uses the correct ultraspherical Bessel function for an open
universe.  Also, in an open universe with curvature constant
\begin{equation}
\label{curv}
  K=H_0^2(\Omega_m+\Omega_v-1)\ ,
\end{equation}
the Poisson equation (\ref{poisson}) is modified; $\nabla^2$ is replaced
by $\nabla^2+3K$ \cite{blh95}.  I define the spatial wavenumber $k$ so that
the eigenfunctions of the Laplacian have eigenvalues $-k^2+K$; $k$ then
has the continuous spectrum $0\le k<\infty$.  Most other workers define
$k$ differently, so that it starts at $\sqrt{-K}$ rather than at 0.  This
is entirely a matter of convention.  However, a power law spectrum with
one choice obviously is not a power-law spectrum with the other one.
Somewhat arbitrarily, {\tt grafic} assumes $P(k)\propto k^{n-4}$ with
$k$ ranging from 0 to $\infty$.  Users interested in open models are
invited to use their own preferred power spectra.

{\tt Grafic} does the numerical integration of the quadrupole anisotropy
by nested quadrature; $\Delta_2(k)$ requires an integral over $\tau$
for each $k$ (cf. eq. \ref{z0corr}), and $Q_{\rm rms-PS}$ then requires
an integration over $k$ (eq. \ref{norm}).  (In case 1 with {\tt linger.dat},
if {\tt linger} is not evolved to $a=1$, then a double quadrature is also
used to correct the results to $a=1$.)  This quadrature takes a few minutes
on a typical workstation; a progress counter is output by {\tt grafic} so
that the user knows it is working.  Romberg integration is used for the
quadratures in {\tt grafic}.  A very small tolerance level is set, which
the integrator sometimes cannot satisfy.  It then prints out a message
``{\tt Rombint failed to converge}....''  Do not worry about this unless
the {\tt error} that follows is larger than $10^{-4}$.

For description of the other input, see section \ref{grafic1}.

\subsubsection{GRAFIC Input 3: Scale-free spectrum}

If the user runs {\tt grafic} with {\tt Tflag = 3}, a transfer function
is not used and neither is the normalization by $Q_{\rm rms-PS}$ and/or
$\sigma_8$.  However, {\tt grafic} still needs to know the cosmological
model parameters, so it prompts the user to input $\Omega_m$, $\Omega_v$,
and $H_0$.  In the scale-free case, one should set $H_0=1$.  After
these values are input, {\tt grafic} requests the long-wave spectral
index $n$; the difference with cases 1 and 2 is that now the present-day
linear spectrum of the potential is exactly $P\propto k^{n-4}$, with no
correction by a transfer function.  {\tt Grafic} assumes that the spatial
curvature scale is sufficiently large so that the corrections to the Laplacian
mentioned in section \ref{grafic2} are negligible.  (If they are not, then
the user had better be performing the nonlinear calculations in a hyperbolic
space!)

The normalization is simpler in the scale-free case than when a transfer
function is used.  {\tt Grafic} sets the normalization at $a=1$ by the
value of $k^3P_\epsilon(k)$ at the shortest wavelength accessible on a
regular lattice, given by the Nyquist frequency $\pi/\Delta x$.   Once
this is set, {\tt grafic} skips directly to the output stage, prompting
the user for {\tt dx = } $\Delta x$ (this should be set to 1), the
force softening length {\tt epsilon}, and the p3m timestep parameter
{\tt etat}, as discussed at the end of section \ref{grafic1}.

\subsubsection{GRAFIC Output}

The output produced by {\tt grafic} is straightforward.  The normalization
values and statistics of the random density and displacement fields are
printed on standard output, while there are two unformatted files giving the
density and particle data, and (optionally) one formatted file giving the
$a=1$ linear matter power spectrum used.  If a large grid size is set in
{\tt grafic.inc}, it may be some time between the last input ({\tt ido},
determining whether an unconstrained or constrained field is to be produced)
and the output of the statistics.  The statistics for the unconstrained
field include ensemble-average values of the rms density contrast and
displacement ({\tt mean sigma\_delta, sigma\_psi}, the latter in units of
comoving Mpc).  The $\chi^2$ statistic gives the sum of squares of the
standard normal deviates (unit-variance, zero-mean gaussian random variables)
generated for the random density field.  It may be compared with its mean
value, {\tt dof} (which equals the number of grid points minus one; one degree
of freedom is eliminated because the density fluctuation field has zero
mean).  {\tt Grafic} also outputs a standardized deviation $\nu$ indicating
by how much $\chi^2$ deviates from its mean value.  Note that $\nu$ for
the unconstrained field is {\it not} related to the magnitude of any imposed
constraint.

If the user applies constraints, statistics are printed out for each
constraint next.  First are the values of the constraint and a
suitably defined $\chi^2$ for the unconstrained field  (for one
constraint of standardized value $\nu_c$, it is $2\nu_c^2$); the
``sampled'' and ``desired'' values differ because the constraint has not
yet been imposed.  {\tt Grafic} indicates when it begins to compute a
constrained realization.  When it is finished, it prints the values of
the constraints computed from the actual density field.  They should
match very precisely the constraints imposed by the user in subroutine
{\tt constr.f}.

The penultimate set of statistics printed by {\tt grafic} are the rms and
maximum density fluctuation and displacement, and $\chi^2$ and the number
of degrees of freedom, for the final random realization at expansion
factor $a=1$ (before the fields are rescaled back to the linear regime).
The rms values should be close (but not equal) to the ensemble-average
values printed out earlier; the maximum values are, of course, several
times larger.  (A 5-standard deviation value is not uncommon on even a
$32^3$ grid.)  If no constraints are imposed, $\chi^2$ and {\tt dof} given
here agree with those for the unconstrained realization; otherwise $\chi^2$
is reduced and {\tt dof} is decreased by the number of constraints.

Finally, {\tt grafic} outputs the expansion factor {\tt astart} to which
it rescales the density fluctuation, displacements, and velocities.
Because the rescaling is done so that the maximum $\delta\rho/\rho=1$,
{\tt astart} is related to the reciprocal of the maximum $\delta\rho/\rho$
printed out for $a=1$.  (In an Einstein-de Sitter universe {\tt astart}
exactly equals the reciprocal because linear density fluctuations grow in
proportion to $a$; in other models the linear growth factor is computed to
give the correct starting time.)  The statistics for the density fluctuation
and displacement fields are then extrapolated back to {\tt astart}.  Most
users will be concerned only with this final set of statistics.

If the user gives the appropriate input, {\tt grafic} produces an ascii
file {\tt power.dat} giving the linear matter power spectrum at $a=1$.
The first line contains the spectral slope $n$ and normalization constant
input by the user; the latter is negative if the normalization is based
on $\sigma_8$ and positive if it is set by $Q_{\rm rms-PS}$.  After
this header follow 201 lines of $(k,P_\epsilon)$, with $k$ logarithmically
sampled from the minimum and maximum values entered by the user.  Note:
$k$ has units of Mpc$^{-1}$ and the power spectrum has units of Mpc$^3$.
COSMICS does not use units of $h^{-1}$ Mpc for length, nor does it use
improperly defined power spectra.  $P(k)$ is a spectral density; it must
be multiplied by a $k$-space volume element to give the power.   Some
experts define $P$ so that the power is $(2\pi)^{-3}P(k)d^3k$; the
absence of factors $(2\pi)^{-3}$ from equation (\ref{norm}) shows
that COSMICS is based on the power being $P(k)d^3k$.

{\tt Grafic} produces two unformatted (binary) files containing the density
fluctuation field and deformed lattice positions and velocities at
expansion factor {\tt astart}.  These files provide the input needed for
nonlinear evolution codes.  The first file, {\tt delta.dat}, contains
$\delta\rho/\rho$ on the lattice and is written as follows:
\hfil\newline{\tt
\indent\indent write(10) np1,np2,np3,dx,astart,omegam,omegav,H0\hfil\newline
\indent\indent write(11) (((delta(i,j,k),i=1,np1),j=1,np2),k=1,np3)
        \hfil\newline
}
(Actually, in the program it is written as a one-dimensional array, but
that is equivalent to the three-dimensional array as shown above.)
The lattice size is ({\tt np1},{\tt np2},{\tt np3}); these numbers
are set in {\tt grafic.inc} before {\tt grafic} is built.

The final file, {\tt p3m.dat}, gives $(\vec x,\vec v\,)$ on the lattice
(see eq. \ref{zeld}) at expansion factor {\tt astart}.  The units of
$\vec x$ are comoving Mpc; those of $\vec v$ are proper km s$^{-1}$.
They are written as one-dimensional arrays because it is more natural to
think of a particle list as being one-dimensional:
\hfil\newline{\tt
\indent\indent write(10) npart,np1,np2,np3,dx,epsilon,astart,omegam,omegav,H0,
        \hfil\newline
\indent\indent \& \indent dt2,etat,nstep,ekin,egrav,egint,nrec\hfil\newline
\indent\indent write(11) ((x(i,j),i=1,3),j=1,npart)\hfil\newline
\indent\indent write(11) ((v(i,j),i=1,3),j=1,npart)\hfil\newline
}
Most of the parameters in the two header lines have been discussed already
(note that {\tt dx} and {\tt epsilon} have units of comoving Mpc, and
{\tt H0} has the natural units); of those that have not, the most
important are {\tt npart} = {\tt np1*np2*np3} (the number of particles)
and {\tt nrec}, which determines how many records are to be used for
writing the positions and velocities.  In most cases, the user will want
to set {\tt nrec} = 1 in {\tt grafic.inc}, in which case {\tt x} and
{\tt v} are written with one record each as shown above.  However, for
very large {\tt npart} and computers with inefficient use of I/O buffers,
it may be difficult or impossible to write {\tt 3*npart} floating-point
numbers as one record.  In that case, increase {\tt nrec}, and examine
the code to see how to read the data back again.  The parameters that
we have not discussed are specific to the author's p3m code.

\section*{Acknowledgments}

I am grateful for the assistance of my collaborators in putting together
this package.  Chung-Pei Ma wrote early versions of the {\tt linger} codes,
based on a crude synchronous-gauge code I had developed previously.
Paul Bode parallelized the {\tt linger} codes, tested them and fixed
numerous problems, and he helped me prepare the makefiles.
Rien van de Weygaert independently wrote a constrained gaussian random
field code and guided my thinking about my own version.  I also thank
Paul Steinhardt and the COMBA collaboration for helpful suggestions,
particularly regarding the importance of photon polarization for CMB
anisotropy computations.

This work was supported by the National Science Foundation under grant
AST-9318185 to the Grand Challenge Cosmology Consortium \cite{gc3}, which
provided the motivation for developing a portable package useful for other
cosmologists.

\def\aa{{Astron. Astrophys.}\ }
\def\araa{{Ann. Rev. Astron. Astrophys.}\ }
\def\apj{{Astrophys. J.}\ }
\def\apjl{{Astrophys. J. (Lett.)}\ }
\def\apjs{{Astrophys. J. Suppl.}\ }
\def\mnras{{Mon. Not. R. Astron. Soc.}\ }
\def\nature{{Nature}\ }
\def\prd{{Phys. Rev. D}\ }
\def\prl{{Phys. Rev. Lett.}\ }
\def\rmp{{Rev. Mod. Phys.}\ }

\end{document}